\newif\ifdraft\drafttrue
\documentclass[accepted]{uai2021} 

\usepackage[american]{babel}

\usepackage{natbib} 
\usepackage{mathtools} 
\usepackage{booktabs} 
\usepackage{tikz} 

\usepackage{url}
\usepackage{graphicx}
\usepackage{booktabs}
\usepackage{algorithm} 
\usepackage{algorithmic}  
\usepackage[algo2e,ruled,linesnumbered,noend]{algorithm2e} 
\usepackage{amsfonts,amssymb,amsmath}
\usepackage{graphicx}
\usepackage{subfigure}
\usepackage{enumitem}
\usepackage{stfloats}
\usepackage{xcolor}
\usepackage{multirow}
\usepackage{breakcites}
\usepackage{titling}

\newtheorem{theorem}{Theorem}

\newtheorem{lemma}{Lemma}
\newtheorem{proof}{Proof}

\usepackage{xcolor}

\title{Contingency-Aware Influence Maximization: A Reinforcement Learning Approach}

\author[1]{\href{mailto:Haipeng Chen <hpchen@seas.harvard.edu>?Subject=Your UAI 2021 paper}{Haipeng~Chen}{}} 
\author[2]{Wei~Qiu}
\author[1]{Han-Ching~Ou}
\author[2]{Bo~An}
\author[1]{Milind~Tambe}
\affil[1]{%
    Center for Research on Computation and Society\\
    Harvard University
}
\affil[2]{%
    School of Computer Science and Engineering\\
    Nanyang Technological University
}

\begin{document}
\maketitle

\begin{abstract}
The influence maximization (IM) problem aims at finding a subset of seed nodes in a social network that maximize the spread of influence. In this study, we focus on a sub-class of IM problems, where whether the nodes are willing to be the seeds when being invited is uncertain, called \textit{contingency-aware IM}. Such contingency aware IM is critical for applications for non-profit organizations in low resource communities (e.g., spreading awareness of disease prevention). Despite the initial success, a major practical obstacle in promoting the solutions to more communities is the tremendous runtime of the greedy algorithms and the lack of high performance computing (HPC) for the non-profits in the field -- whenever there is a new social network, the non-profits usually do not have the HPCs to recalculate the solutions. Motivated by this and inspired by the line of works that use reinforcement learning (RL) to address combinatorial optimization on graphs, we formalize the problem as a Markov Decision Process (MDP), and use RL to learn an IM policy over historically seen networks, and generalize to unseen networks with negligible runtime at test phase. To fully exploit the properties of our targeted problem, we propose two technical innovations that improve the existing methods, including state-abstraction and theoretically grounded reward shaping. Empirical results show that our method achieves influence as high as the state-of-the-art methods for contingency-aware IM, while having negligible runtime at test phase.
\end{abstract}

\section{Introduction}\label{sec:intro}
Influence maximization is the problem of finding a subset of seed nodes in a social network that maximize the spread of influence. Originally derived from the viral marketing domain, the majority of IM algorithms~\citep{kempe2003maximizing,leskovec2007cost,borgs2014maximizing,tang2015influence} focus on settings where nodes are always willing to be the seeds, which may not be the case in many real-world scenarios. For example, recent work~\citep{yadav2016using,yadav2018please,wilder2018end} provides large-scale applications of IM in public health. More specifically, they  use IM algorithms to help spread the awareness of HIV prevention among homeless youth, where the youth leaders when being invited to be the ``seed'' nodes, may have difficulty joining and thus deviate from the intervention plan. 
In this sub-class of IM problems called \textit{contingency-aware influence maximization}~\citep{yadav2018please}, when a node is invited to become a seed node, there is \textit{uncertainty} in whether it is willing to accept the invitation.

\begin{figure}
  \begin{minipage}[c]{0.62\columnwidth}
    \includegraphics[width=\columnwidth]{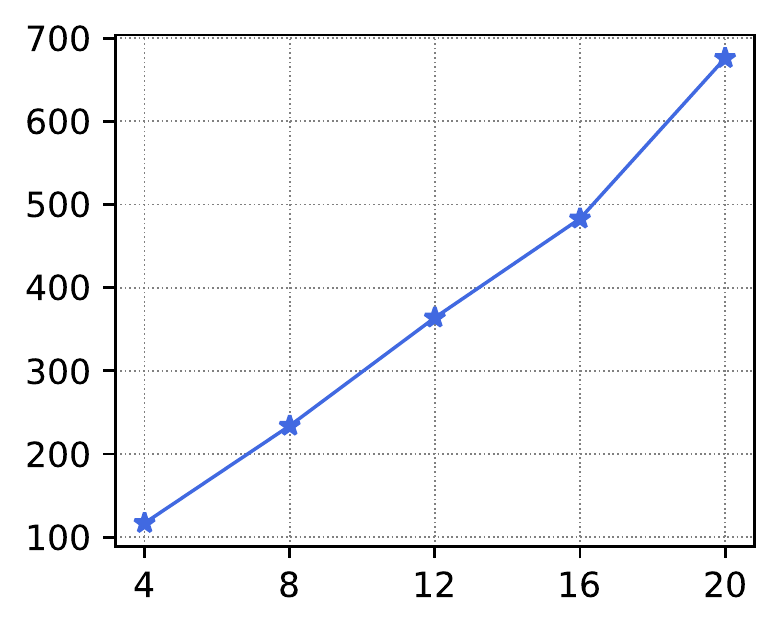}
  \end{minipage}\hfill
  \begin{minipage}[c]{0.35\columnwidth}
    \caption{
      Runtime of CHANGE (in mins), where x-axis is the number of seed nodes.
    } \label{fig:change_runtime}
  \end{minipage}
\end{figure}

This contigency-aware IM problem has been addressed using 
Partially observable MDP (POMDP)~\citep{yadav2016using,yadav2018please} and greedy algorithms \citep{wilder2018end}. Despite their success in the field~\citep{wilder2021clinical}, there is a major limitation in transitioning the solution to more homeless youth shelters and cities -- whenever the underlying social network changes, the solution to the IM problem needs to be recomputed, whereas the stakeholders usually do not have the HPCs to perform the computation on their own. Figure~\ref{fig:change_runtime} shows the runtime of the IM component for the state-of-the-art CHANGE algorithm~\citep{wilder2018end,wilder2021clinical} on a network with 1000 nodes and 4974 edges when the number of seeds increases (run on a single Intel(R) Core(TM) i9-9820X CPU @ 3.30GHz core, with the sampling frequency of CHANGE set to 50, the influence propagation probability being set to 0.2). CHANGE runs 10 hours even with just 20 seed nodes, presenting a great burden to the low-resource non-profits such as homeless shelters, particularly as they scale-up these applications.
In fact, the low-resource computing issue exists in many other works on social network intervention~\citep{srivastava2019network,rice2020using,awasthi2020learning,petering2021examining} and deploying AI techniques to the public sector in general~\citep{mehr2017artificial,mikhaylov2018artificial,guo2018application}, especially when low-resource non-profits are the decision makers at stake. 

Recently, there have been studies that use RL to learn a generalized policy for a certain combinatorial optimization problems on graphs~\citep{khalil2017learning,nazari2018reinforcement,deudon2018learning,bengio2020machine}. The key idea is to decompose the selection of nodes into a sequence, and learn a heuristic policy that selects nodes sequentially. The RL policy is usually trained on a set of seen training graphs, in the hope that it generalizes to unseen test graphs of similar characteristics. To better generalize the trained policy across different graphs, graph embedding techniques, such as Structure to Vector (S2V)~\citep{dai2016discriminative} and Graph Convolutional Networks (GCNs)~\citep{kipf2016semi} are integrated as part of the RL value functions to extract the graph structure information.
Primarily proposed to solve relatively simple problems such as the traveling salesman problem (TSP) and the maximum vertex cover (MVC) problem, recent works~\citep{li2019disco,manchanda2020gcomb,tian2020deep} extend it to the IM problem without considering node uncertainty. 
Inspired by these works, we propose to address the contingency-aware IM problem using RL.


There are however new challenges in designing an effective RL algorithm for contigency aware IM. 
First, in previous RL for IM methods, the state (as well as state transition) of the MDP is the nodes that are previously selected, which is deterministic. In our problem, the willingness status of nodes selected before the current step are unknown. When formulating an MDP model, there remains a question of how to define a state that well incorporates the uncertainty information. Second, in these previous works, the immediate reward is set as the marginal contribution of a new node selected at the current time step. This cannot be simply applied in our problem because of the uncertainty in node status. Moreover, it introduces an extremely high variance in the marginal contribution of a new node, and thus renders the RL training much more challenging.

To address the challenges, we propose a new MDP formulation to the underlying problem. Though this formulation preserves the Markovian property, its state is highly sparse and thus makes it hard for RL to learn efficiently. We make the first technical innovation with a state-abstraction component for RL, which compresses the states in a more compact manner, while preserving the uncertainty information. To address the high variance in reward function, we make our second technical innovation by using a novel reward-shaping technique. The reward-shaping component exploits two unique properties in the problem: 1) The probability of a generic node willing to be a seed or not is usually known, which can be learned from historical data; 2) The influence function is submodular. We first use the node willingness probability to express the exact ``expected'' reward, which we show is computationally infeasible. Using the submodularity property, we then design a surrogate reward function in place of the exact expected reward, with provable worst-case guarantee compared to the exact expected reward.


\noindent\textbf{Summary of contributions.}
1) We are the first to address the contingency-aware IM problem using an RL approach. We propose a new MDP formulation for this problem. 2) Our technical contribution is a new RL algorithm that is built upon the line of works that use RL to address combinatorial optimization problems on graphs, while making non-trivial, theoretically grounded adaptations that exploit the problem properties. 
3) We conduct extensive experimental evaluations and show that under various settings, RL can perform as good as state-of-the-art greedy IM algorithms from the HIV prevention domain. Ablation study results demonstrate the effectiveness for each of the two novel components. Our code can be found via {\url{https://github.com/Haipeng-Chen/RL4IM-Contingency}}.

\section{Related Work}

\noindent\textbf{Influence maximization} 
The IM problem is first studied by~\cite{domingos2001mining} as an algorithmic problem. \cite{kempe2003maximizing} formulate it as a discrete optimization problem over the graphs, and propose a greedy algorithm to solve the problem, which has a guarantee of $1-1/e$.  
Cost-Effective Lazy Forward (CELF)~\citep{leskovec2007cost}, Reverse Influence Sampling (RIS)~\citep{borgs2014maximizing} and Influence Maximization via Martingales (IMM) improve the greedy algorithm by more efficient spread estimation techniques. \citet{golovin2011adaptive} extend the IM problem to the \textit{adaptive} setting, where seed selection is adapted based on observing the influence spread of previously selected nodes. 
More efficient methods~\citep{han2018efficient,sun2018multi,huang2020efficient} are proposed later on to solve the adaptive IM problem. Adaptive IM differs from our setting in that   our focus is to address the \textit{uncertainty} in a node's willingness to participate, which they do not address. Moreover, they observe the influence of the previously selected nodes and select new nodes based on the observation, whereas we do not observe such intermediate influence. 
\citet{yadav2018please} introduce contigency-aware influence maximization in the context of HIV prevention among homeless youth and solve the challenge using a POMDP; this solution does not scale beyond very small number of influencers. To remedy this shortcoming, \citet{wilder2018end} develop greedy IM algorithms for contingency aware influence maximization in the field, which is deployed in field test~\citep{wilder2021clinical}. Different from their work, we introduce learning techniques to address the problem.

\noindent\textbf{ML/RL for combinatorial optimization on Graphs}
\citet{vinyals2015pointer,bello2016neural,graves2016hybrid} make early attempts in using ML/RL to address combinatorial optimization problems on graphs, where they decompose the original combinatorial action into a sequence of individual actions, and propose learning frameworks to learn heuristics for the problems. 
These approaches do not generalize well among unseen graphs, or are data-inefficient. \citet{khalil2017learning} propose to use graph embedding techniques as the value approximator for the Deep Q-Networks (DQN)~\citep{mnih2013playing}, and therefore their approach generalizes better for graphs out of distribution. 
\citet{kool2018attention} propose an approach that combines attention-based function approximators with policy gradient methods~\citep{williams1992simple}. \citet{li2018combinatorial} approximate the solution quality with GCNs~\citep{kipf2016semi}, and use a learning framework based on guided tree search. \cite{joshi2019efficient} address the problem using a combination of GCNs and beam search. \citet{qiu2019dynamic} combine RL and GCNs to address the road tolling problem in a transportation network. \citet{ou2021active} adapt the idea to address recurrent disease prevention on a social network. \citet{mao2019learning} use it to address the scheduling problem in data processing clusters. We refer to \cite{bengio2020machine} for a detailed survey on this line of works.

\noindent\textbf{ML/RL for IM} 
\citet{lin2015learning,ali2018boosting} use RL to do influence maximizaiton in a competitive setting. They do not consider generalization, and the policy is to choose which high-level greedy algorithm to use. \citet{kamarthi2020influence} apply RL to explore an unknown graph in the context of influence maximization. This work is different from ours as they use RL to explore the graph structure instead of selecting seeds. \citet{ko2020monstor} propose an inductive ML approach to estimate the influence spread of unseen networks. \citet{li2019disco,tian2020deep,manchanda2020gcomb} extend the method in~\citep{khalil2017learning} to address the IM problem, where reward of a new node is defined as its marginal contribution.
\citet{manchanda2020gcomb} aim at solving problem instances with millions of nodes. It uses supervised learning as a preliminary step to predict the individual quality of a node, which introduces large extra computational overhead and effort of hand-crafting the learning pipeline. Because of this, it does not scale to large number of training graphs. Moreover, all these methods do not consider the uncertainty of a node's willingness to be seed, and thus fail to address the challenges that are discussed previously. We will show empirically that directly applying their methods leads to sub-optimal performance.

\section{Contingency-Aware IM }\label{sec:model}

Our work is motivated by previous works~\citep{yadav2016using,yadav2018please,wilder2018end} which use influence maximization to spread the awareness of HIV prevention among the homeless youth. An HIV awareness intervention is a day-long class followed by weekly hour-long meetings. Due to limited resources, only a subset of youth will be selected to attend the classes. The trained youth will then act as peer-leaders who further spread the awareness of HIV prevention among the youth social network. An important observation is that when being invited to the training sessions, whether the youth is willing to be present at the classes is unknown until the end of the intervention round. 
Despite the initial success, a practical challenge that prevents the transitioning of their methods to more homeless youth shelters and cities is the lack of HPC resources for the non-profits in the field. Once the social network changes, the algorithm has to be rerun, without reusing knowledge about historical data. In fact, the \textit{low-resource computing} scenario is ubiquitous in real life, especially for low-resource non-profits which are in urgent need of help from the AI community. We provide a new learning-based perspective of addressing the problem.

\subsection{Problem Setup}

\noindent\textbf{Influence spread model} We consider a social network $G=(V,E)$ where $V$ and $E$ are respectively the nodes and edges. 
Each node is either \textit{activated}, meaning the node is influenced, or \textit{inactivated} otherwise. We assume all nodes are initially inactivated unless chosen as the seed node. Two nodes that are connected by an edge $e\in E$ has a probability of influencing each other.
We model the influence spread using the prominent Independent Cascade (IC) model~\citep{goldenberg2001talk,kempe2003maximizing}. That is, for each node $v\in V$ that is activated at a certain time, it has a \textit{single} chance of activating its neighbors at the next time, with a probability $p$. Given a seed node set $S\subseteq V$, the influence spread in the graph $G$ is represented as $I(G,S)$. The \textit{influence maximization} problem aims at finding an optimal set of (usually budget-constrained) seed nodes $S^*\subseteq G$, such that the influence spread is maximized.

\noindent\textbf{Seed selection and node uncertainty}
As motivated by the multi-round seed selection in the HIV prevention domain, we consider seed selection as multiple rounds $t=1\ldots T$ of seed nodes selection, where each round selects a mini-batch of $B$ nodes.\footnote{Note that our model is not limited to the multi-round setting, but is a more generalized model that can tackle both single round and multi-round seed selections. We will show empirically that our model and algorithm work well on single round node selection.} At each round $t$, the set of selected seed nodes is represented as ${S}_t$. As discussed before, when being selected, each node $v\in S_t$ may not necessarily be willing to act as a seed node. 
To capture this uncertainty, we denote the probability of a node willing to be seed (when selected as seed node) as $q$.\footnote{Similar to~\cite{yadav2018please}, we assume a same $q$ value for all the nodes in this model due to the practical challenge of knowing the exact $q$ value for each node. In~\cite{yadav2018please}, it is done by using statistics on the historical attendance rate of youths when being invited.} We assume that this probability is known \textit{a priori}, which can be estimated by using statistics of historical data.
The realization of the willingness status of the selected nodes can be observed at the end of each intervention round $t$. Naturally, the set of nodes $O_t$ who are willing to be seeds at round $t$ is a subset of $S_t$: $O_t\subseteq S_t$. At the end of $t$, the history of selection and willingness status of all seeds is denoted as a sequence ${H}_t = (({S}_1, {O}_1)\ldots ({S}_t, {O}_t))$.

\subsection{MDP Formulation}\label{sec:mdp}

Due to the sequential planning essence of the multi-round IM problem, we formulate it as a discrete time MDP. 

\noindent\textbf{Time step} A natural way of defining a time step is to treat each intervention round $t$ as a time step. However, in doing so, the action of each time step still consists of $B$ nodes, and thus selecting the optimal action in each round $t$ is still a combinatorial optimization problem with a combination of choices of size $|V| \choose B$. To avoid this, we define the time step as selecting each individual node. To distinguish the two concepts, we will call each intervention round as a \textit{main step} $t$, and the selection of each individual node within each main step as a \textit{sub-step} $(t,b)$. We have $t=1\ldots T$, and $b=1\ldots B$. The time horizon is thus $T\times B$. 

\noindent\textbf{State} To fully capture the information of the status of a current sub-step $(t,b)$, we use a binary matrix $X_{t,b}\in \{0,1\}^{3\times |V|}$, together with the adjacency matrix $G$ to represent the state $(G,X_{t,b})$.\footnote{With a bit abuse of notation, we use $G$ to represent both a graph and its adjacency matrix.} Note that $G$ is fixed over time. Thus we will just use $X_{t,b}$ to refer to the state. Each column $X_{t,b}^v$ of $X_{t,b}$ denotes the status of one node $v$. In the initial state, $X_{t,b}$ is initialized as all zeros. As the sequence of decision goes, the first element $X_{t,b}^{1,v}=1$ indicates node $v$ is selected as a seed node and is willing to be seed. The second element $X_{t,b}^{2,v}=1$ means node $v$ is selected as a seed node and is unwilling to be seed. The third element $X_{t,b}^{3,v}=1$ means node $v$ is selected at a main step $t$ but the main step is not ended, so that its willingness status remains unknown. In this way, we can compress the history $H_t$ of node selection and realization of nodes' willingness status using a matrix form. Given the Markovian property, the status of the current time step does not depend on the sequence. Moreover, it considers the uncertainty in nodes' willingness status within the current main step. Thus, the state representation does not lose information about the state.



\noindent\textbf{Action} There are two types of actions. We define a \textit{sub action} as the selection of a single node at each sub-step $(t,b)$. It is denoted as a one-hot vector $a_{t,b}\in \{0,1\}^{|V|}$, where there is only one element in $a_{t,b}$ that corresponds to the node being selected, i.e., $\sum_{v=1}^{|V|} a_{t,b}^v=1$. Correspondingly, a \textit{main action} $A_t$ is defined as the aggregation of all the sub actions in this main step at the end of each main step $t$:
\begin{align}
    A_t = \sum\nolimits_{b=1}^B a_{t,b}, \ \forall t=1 \ldots T
\end{align}
$A_t$ can be seen as the vector form representation of $S_t$. Note that $\sum_{v=1}^{|V|}A_t^v=B$.

\noindent\textbf{State transition} We omit the description of $G$ as it is fixed over time. Apart from that, there are two types of state transitions. The first type happens at the end of each sub-step $(t,b)$, i.e., whenever a new node is selected, there is:
\begin{align}\label{eq:state_trans1}
    X_{t,b+1}^{3} = X_{t,b}^{3}+a_{t,b},\  \forall b=1\ldots B-1, \ t=1\ldots T
\end{align}
We can see that this type of state transition is deterministic. 

The second type of sate transition happens only at the end of each main step $t$. It reveals the realization of the willingness status of the selected nodes at the main step. This type of state transition is stochastic, and directly depends on the probability $q$. To formalize it, we first define $\bar{A}_t\in \{0,1\}^{|V|}$ as the realization of main action $A_t$, which can be seen as the vector form of $O_t$. The $v$-th element $\bar{A}_t^v=1$ means that node $v$ is invited and willing to be a seed. It is naturally constrained that for each $v\in V: \bar{A}_t^v\leq A_t^v$. Let a scalar $\bar{B}_t\coloneqq \sum_{v=1}^{|V|}A_t^v$. It means the number of nodes which are willing to be seeds when being selected at main step $t$. Given the above, we can derive the three dimensions of state $X_{t+1,1}$ of the next time step $(t+1,b=1)$ as 
\begin{align}
\begin{aligned}\label{eq:state_trans2}
&X_{t+1,b=1}^1 = X_{t,B}^1+\bar{A}_t, \\
&X_{t+1,b=1}^2 = X_{t,B}^2+A_t-\bar{A}_t,\ \ 
X_{t+1,b=1}^3 = 0
\end{aligned}
\end{align}
The probability of this transition is:
\begin{align}\label{eq:state_trans3}
    P(X_{t+1,1},A_t\rightarrow \bar{A}_t|X_{t,B},A_t)= q^{\bar{B}_t}(1-q)^{B-\bar{B}_t}
\end{align}

\noindent\textbf{Reward} The total reward is defined as the total influence that is achieved within the social network $G$, given the selection of nodes that is represented as $X_{T,B}$. We denote the total accumulated reward as $r(G,X_{T,B})$. This incurs the issue known as reward sparseness, which makes it challenging for RL to learn efficiently. To mitigate this issue, \citet{li2019disco,tian2020deep,manchanda2020gcomb} use the marginal contribution of a new node as the \textit{immediate reward}. Denote the set of seed nodes selected before as $S$, the \textit{marginal contribution} of a new node $v$ is defined as as $\Delta I(G,S,v) \coloneqq I(G,S\cup \{v\})-I(G,S)$. This cannot be directly applied to our problem due to the node willingness uncertainty within each intervention round $t$. For now we denote the immediate reward as a generic notation $r(G,X_{t,b}, a_{t,b})$. We will revisit this issue in Section~\ref{sec:reward_shaping} when we introduce our proposed reward shaping technique.

\section{RL4IM}
Inspired by~\citep{khalil2017learning,nazari2018reinforcement,deudon2018learning,bengio2020machine} that use RL and graph embedding to address combinatorial optimization problem on graphs, we design Reinforcement Learning for Influence Maximization (RL4IM), a new RL-based algorithm that addresses the contingency-aware IM problem. RL4IM exploits two significant properties in the underlying problem: 1) The influence function is submodular; 2) The state transition probability, i.e., the probability $q$ of a node willing to be a seed is known \textit{a priori}, which can be estimated using historical data. 

\begin{figure*}[ht]
    \centering
    \includegraphics[width=0.75\linewidth]{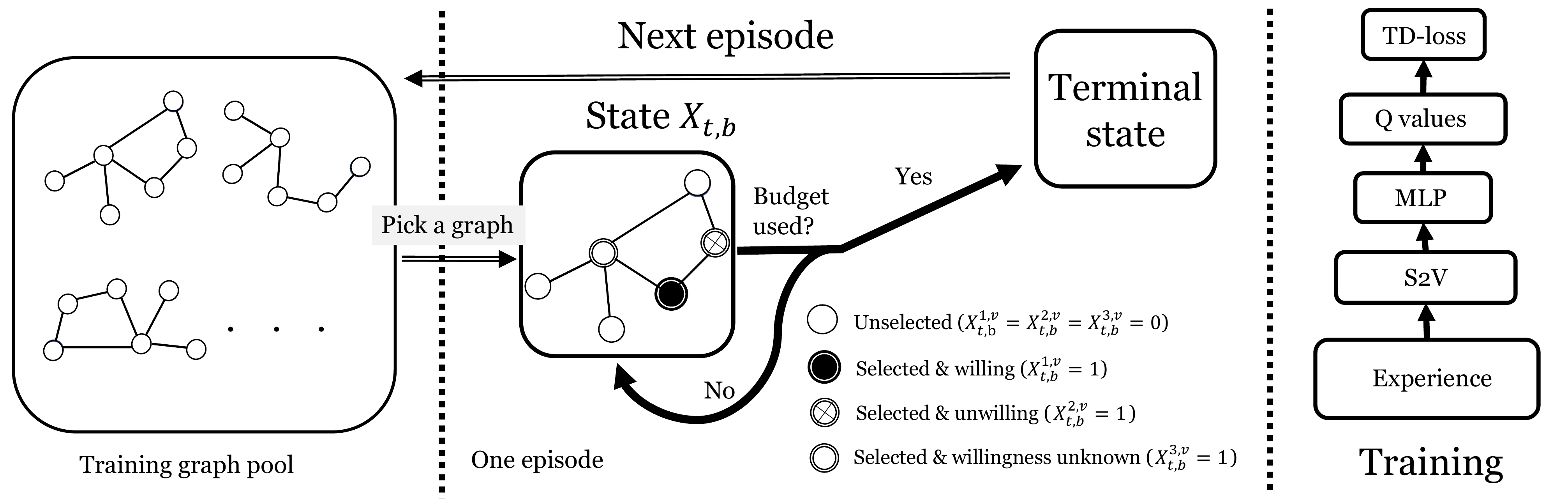}
    \caption{RL4IM training procedure. The graphs on the left are the set of training graphs $\mathcal{G}$. The process starts by randomly selecting a graph $g\in \mathcal{G}$. The sampled graph constructs a new environment. The RL4IM agent then interacts with it. At each time step, it observes the state from the environment, and selects the next node (based on its Q-function) if the budget is not used, or otherwise reaches the terminal state. It then selects the next graph and the training iterates. Meanwhile, the trajectory data are fed into the replay buffer, which is then used to compute the TD loss for updating the Q-function (on the right).}
    \label{fig:architecture}
\end{figure*}

\subsection{RL4IM Architecture}

Figure~\ref{fig:architecture} shows the overall architecture of RL4IM. The graphs on the left are the set of training graphs $\mathcal{G}$. The process starts by randomly selecting a graph $g$ from the set of training graphs. Each sampled training graph constructs an \textit{environment}, which defines a new MDP as described in Section~\ref{sec:mdp}. Given the environment, the RL4IM agent then interacts with it in discrete time steps. At each time step $(t,b)$, it observes the \textit{state} $X_{t,b}$ from the environment, and decides whether the budget of selecting the seed nodes (i.e., $T\times B$) is spent. If not, the RL4IM agent will determine the next seed to select based on its learned \textit{policy} $\pi(a_{t,b}|X_{t,b})$, which is a probability distribution over the feasible action space given the current state $X_{t,b}$ that trades off \textit{exploiting} a node with an estimated high reward and \textit{exploring} nodes that could potentially have higher reward. If the budget is spent, then it reaches the \textit{terminal state} for this \textit{episode}. The RL4IM agent will then select the next graph and the training procedure iterates until its policy reaches convergence. 

In Q-learning~\citep{watkins1989learning}, the value of a node/action is measured using the \textit{Q-function}. The Q-function is usually estimated using the Bellman equation: 
$Q(X_{t,b},a_{t,b}) = r(X_{t,b},a_{t,b})+\gamma \arg\max\nolimits_{a_{t',b'}}Q(X_{t',b'},a_{t',b'})$, where $\gamma$ is the discount factor. 
DQN~\citep{mnih2013playing,mnih2015human} improves vanilla Q-learning by using deep neural networks as the \textit{function approximator}, along with other techniques like experience replay, which stores the historical training trajectories in a \textit{replay buffer} $\mathcal{M}$ and updates the Q-function by minimizing the loss function $(\hat{Q}-Q)^{2}$ with batch data from the replay buffer using gradient descent.

We follow \citet{khalil2017learning,li2018combinatorial} and generalize the learned policies to unseen test graphs using graph/node embedding techniques~\citep{dai2016discriminative,kipf2016semi} as the \textit{function approximator}. Essentially, graph/node embedding takes input an attributed matrix (the state $X_{t,b}$ and the action $a_{t,b}$ in our case) as well as the adjacency matrix $G$ and maps it to an embedding space. It aggregates the neighborhood information from the adjacency matrix. We omit their explicit form, and represent them with a generic form as $f(X_{t,b},G)$ and $g(a_{t,b},G)$. In this way, the Q-function is represented as: $Q(X_{t,b},a_{t,b})=MLP(f(X_{t,b},G),g(a_{t,b},G))$, where $MLP(\cdot)$ means multi-layer perceptron. 
Alg. \ref{alg:rl4im} shows the pseudo-code of RL4IM's training process, where the key novel components are state-abstraction and reward shaping.

\begin{algorithm2e}[htb]
\small
\SetAlgoNoEnd
\setcounter{AlgoLine}{0}
Initialize replay buffer $\mathcal{M}$, Q-function $Q_{\theta}(\bar{X}_{t,b},a_{t,b})$

\For{episode: 1 to \#episodes}{
Draw a graph $G\in \mathcal{G}$

Get initial state $X_{1,1}=0$

\For{$t= 1\ldots T$}{
\For{$b=1\ldots B$}{
Get abstracted state $\bar{X}_{t,b}\gets X_{t,b}^1+qX_{t,b}^3$ 

Get action from policy $a_{t,b}\gets \arg\max Q_{\theta}(a_{t,b}|\bar{X}_{t,b})$ with probability $1-\varepsilon$ or otherwise random

Play $a_{t,b}$, get surrogate reward $\tilde{r}(X_{t,b},a_{t,b})$ with Eq.\eqref{eq:surrogate_reward}

\lIf{b<B}{ $X_{t',b'}\gets X_{t,b+1}$ with Eq.\eqref{eq:state_trans1}}
\lElse{$X_{t',b'}\gets X_{t+1,1}$ with Eqs.\eqref{eq:state_trans2}-\eqref{eq:state_trans3}}

Add new memory to replay buffer: $\mathcal{M}=\mathcal{M}\cup (\bar{X}_{t,b}, a_{t,b}, \tilde{r}(X_{t,b},a_{t,b}),  X_{t',b'})$

Update $\theta$ using sampled memories from $\mathcal{M}$
}
}
}

\Return $Q_{\theta}(\bar{X}_{t,b},a_{t,b})$
\caption{RL4IM training}\label{alg:rl4im}
\end{algorithm2e}

\subsection{State abstraction}

We use a $3\times |V|$ matrix to capture all the information of the current state. Because only a small subset of nodes are selected as seed nodes, the number of $1$'s is bounded by the total budget $T\times B$. Moreover, the $3$rd dimension contains the status of nodes at only one intervention round, and is bounded by the budget $B$ at each intervention round. 
Therefore, the state matrix is extremely sparse and makes learning rather inefficient. To address this issue, instead of assuming that we do not know about the state-transition model -- as typical model-free RL methods do -- we exploit the fact that the transition model is actually known. That is, we know the probability $q$ that a node is willing to be seed, which can usually be learned from historical data.\footnote{
We do not know its realization till the main step ends, though.} 
More specifically, we use a more compact vector $\bar{X}_{t,b}\in \mathcal{R}^{1\times |V|}$, that performs a state abstraction to $X_{t,b}$:
\begin{align}
    \bar{X}_{t,b} = X_{t,b}^1 + qX_{t,b}^3
\end{align}
By multiplying the $3$rd dimension with the probability $q$, the intuition is to use this prior knowledge to better reflect the ``expected'' contribution of the corresponding node. Note that in the abstracted state, the information about the nodes that are selected but are unwilling to be seeds are not tracked. To keep track of this information, we maintain a feasible action set that is updated at each time step. The set is initialized as the entire set of nodes of the graph. In every time step, whenever a node is selected, it will be removed from the set so that it is no longer feasible in future time steps.

\subsection{Reward shaping}\label{sec:reward_shaping}

As discussed in Section~\ref{sec:mdp}, to mitigate the reward sparseness issue, existing works~\citep{li2019disco,tian2020deep,manchanda2020gcomb} use the marginal contribution of a newly selected node as the immediate reward at the current time step. However, this is infeasible in our problem when there is uncertainty about the nodes' willingness status in each main step. A straightforward way of handling it is just to assume that all the nodes in the current main step are willing to be seeds, and calculate the marginal contribution with respect to all these nodes.  
However, due to submodularity of the influence function, this incurs underestimation of a new node's marginal contribution. As discussed previously, we have prior knowledge about the node willingness status transition probability $q$. We then use it to explicitly represent the expected marginal contribution of a new node. Recall that $\bar{A}_t$ denotes the realization of the main step action $A_t$. Let $\bar{A}_{t,b}$ be the realization of sub-steps $a_{t,1}$ to $a_{t,b-1}$ (or equivalently $X_{t,b}^3$, as $X_{t,b}^3=a_{t,1}+\ldots+a_{t,b}$), then $\bar{B}_{t,b}\coloneqq \sum_{v=1}^{|V|}\bar{A}_{t,b}$ means the number of nodes that are selected from $(t,1)$ to $(t,b)$ and are willing to be seeds. Thus, the explicit form of expected marginal contribution $r(X_{t,b},a_{t,b})$ of  action $a_{t,b}$ is:
\begin{align}\label{eq:reward}
\hspace{-2mm} \sum\nolimits_{\beta=0}^{b-1} q^{\beta}(1-q)^{b-1-\beta}\sum\nolimits_{X_{t,b}^3}\delta I(G,X_{t,b},a_{t,b})\Big|\bar{B}_{t,b}=\beta,
\end{align} where $\delta I(G,X_{t,b},a_{t,b})=I(G,X_{t,b},a_{t,b})-I(G,X_{t,b})$ is the marginal contribution of action $a_{t,b}$ given the current state $X_{t,b}$, and $\sum_{X_{t,b}^3}\delta I(G,X_{t,b},a_{t,b})$ is the sum of marginal contribution of action $a_{t,b}$ over all possible values of the state's $3$rd dimension $X_{t,b}^3$. The condition $\bar{B}_{t,b}=\beta$ specifies that the number of nodes that are willing to be seeds in the realization of $X_{t,b}^3$.
We can see that there are ${b \choose \beta}$ such terms in the summation. 

In practice, the exact influence values $I(G,X_{t,b},a_{t,b})$ and $I(G,X_{t,b})$ are not known, and must be estimated by running multiple influence spread simulations over the graph $G$. Therefore, it becomes computationally infeasible to enumerate all the possible combinations of $X_{t,b}^3$ at each sub-step, especially when the budget $B$ at each main step is large. This becomes a major obstacle for RL, as it usually requires a large number of training samples (time steps) to learn.

To overcome this obstacle, we notice that the influence function $I(G,S)$ is usually \textit{submodular}~\citep{kempe2003maximizing}, meaning that the marginal contribution of a node $v$ when being added to an existing set of nodes $S$, is no larger than that when it is added to subset $S'\subseteq S$, i.e.,
\begin{align}
    I(G, S\!\cup\!\{v\})\!-\!I(G,S)\! \leq \!I(G,S'\!\cup\! \{v\})\!-\!I(G,S')
\end{align}

At each time step $(t,b)$, we use $\delta I_0$ to denote the marginal contribution of an action/node $a_{t,b}$ when no node selected from $(t,1)$ to $(t,b-1)$ is willing to be a seed, and use $\delta I_{b-1}$ to denote the marginal contribution of an action $a_{t,b}$ when all of these nodes are willing to be seeds. That is,
$
\delta I_0 = I(G,X_{t,b},a_{t,b}) -I(G,X_{t,b}) \Big | \bar{B}_{t,b}=0,
\delta I_{b-1} = I(G,X_{t,b},a_{t,b}) -I(G,X_{t,b}) \Big | \bar{B}_{t,b}=b-1
$.
Following the submodularity property of the influence function, we have
\begin{lemma}\label{lemma:bound}
At sub-step $(t,b)$, the marginal contribution of any action is bounded by $\delta I_0$ and $\delta I_{b-1}$:
\begin{align}
    \delta I_{b-1}\leq \delta I(G,X_{t,b},a_{t,b})\leq \delta I_0
\end{align}
\end{lemma}
\begin{proof}
According to the submorularity property of the influence function, i.e., for any subset $S'\subseteq S$,
\begin{align*}
    I(G, S\cup \{v\})-I(G,S) \leq I(G,S'\cup \{v\})-I(G,S')
\end{align*}
Recall that $O_t$ denotes the set of nodes that are willing to be seeds at round $t$, then $O_1\cup \ldots \cup O_{t-1}$ denotes the set of nodes that are willing to be seeds before $t$. We denote $S^{t,b}$ as the set of nodes that are selected in $t$ before $b$, and $O_{t,b}$ as the set of nodes that are willing to be seeds at round $t$ before $b$. 
Because for any $O_{t,b}$, there is $\emptyset \subseteq O_t\subseteq S_{t,b}$. Therefore $O_1\cup \ldots \cup O_{t-1} \subseteq O_1\cup \ldots \cup O_{t-1} \cup O_{t,b} \subseteq O_1\cup \ldots \cup O_{t-1} \cup S_{t,b}$. By definition we have Lemma 1.
\end{proof}

Using this property, we design a surrogate marginal contribution function of $\delta I(G,X_{t,b},a_{t,b})$, where we assume for any realization $\bar{A}_{t,b}$ of ${X}_{t,b}^3$, the marginal contribution of an action $a_{t,b}$ is the same when $\bar{B}_{t,b}=\beta$, i.e., for any two states $X_{t,b}$ and $X'_{t,b}$ and their corresponding $\bar{B}_{t,b}$ and $\bar{B}'_{t,b}$:
\begin{align}\label{eq:assume1}
\begin{aligned}
&\bar{B}_{t,b}=\bar{B}'_{t,b}\Rightarrow \\
&\delta I(G,X_{t,b},a-{t,b})=\delta I(G,X'_{t,b},a-{t,b})
\end{aligned}
\end{align}
This assumption means the marginal contribution of a new node only depends on \textit{how many} nodes are willing to be seeds, not \textit{which}. We can then denote the marginal contribution as $\delta I_{\beta}$. We futher assume that $(\delta I_0 \ldots \delta I_{\beta} \ldots \delta_{b-1})$ is an arithmetic sequence of common difference, i.e., 
\begin{align}\label{eq:assume2}
    \delta I_{\beta} = \delta I_0 + \beta \Delta,
\end{align} where $\Delta\coloneqq (\delta I_{b-1}-\delta I_0)/(b-1)$ is the common difference. Due to Lemma~\ref{lemma:bound}, $\Delta\leq 0$.

\begin{theorem}\label{thm:surrogate_reward}
With the assumptions in Eqs.\eqref{eq:assume1}-\eqref{eq:assume2}, the surrogate marginal contribution of action $a_{t,b}$ in Eq.\eqref{eq:reward} is:
\begin{align}\label{eq:surrogate_reward}
    \tilde{r}(X_{t,b},a_{t,b})= (1-q)\delta I_0+q \delta I_{b-1}
\end{align}
\end{theorem}
\begin{proof}
By substituting Eqs.\eqref{eq:assume1}-\eqref{eq:assume2} into Eq.\eqref{eq:reward}, we have 
\begin{align*}
\tilde{r}(X_{t,b},a_{t,b})&=\sum_{\beta=0}^{b-1}q^{\beta}(1-q)^{b-1-\beta} {b-1\choose\beta}\delta I_{\beta}\\
&= \sum_{\beta=0}^{b-1}q^{\beta}(1-q)^{b-1-\beta} {b-1\choose \beta} (\delta I_0+\beta \Delta )\\
&= \sum_{\beta=0}^{b-1}q^{\beta}(1-q)^{b-1-\beta} {b-1\choose \beta} \delta I_0 + \\ &\Delta \sum_{\beta=0}^{b-1}q^{\beta}(1-q)^{b-1-\beta} {b-1\choose \beta} \beta\\
&=\delta I_0 + \Delta (b-1)q \\
&=\delta I_0 + q(\delta I_{b-1}-\delta I_0)\\
&=(1-q) \delta I_0+q \delta I_{b-1}
\end{align*}
The 4th equation holds because for arithmetic sequence with common difference, there is $ \sum_{\beta=0}^{b-1}q^{\beta}(1-q)^{b-1-\beta} {b-1\choose \beta}=1$, and $ \sum_{\beta=0}^{b-1}q^{\beta}(1-q)^{b-1-\beta} {b-1\choose \beta} \beta=(b-1)q$.
\end{proof}

Despite the simple form, we have the following two desirable properties of the surrogate reward function.
\begin{theorem} \label{thm:complexity}

Using the surrogate marginal contribution in Eq.\eqref{eq:surrogate_reward}, the computational complexity at each step $(t,b)$ reduces from $\mathcal{O}(2^{b})$ to $\mathcal{O}(1)$.
\end{theorem}
\begin{proof}
Because for each $\bar{B}_{t,b}=\beta$, we need to calculate the marginal contribution $b-1 \choose \beta$ times, the total number of of calculations is then $2\times \sum_{\beta=0}^{b-1} {b-1 \choose \beta}=2^b \sim \mathcal{O}(2^b)$. The number 2 at the LHS of the equation means calculating once for both the minuend and the subtrahend. On the other hand, calculating Eq.\eqref{eq:surrogate_reward} requires only calculating $2\times 2=4$ influence values, which is of order $\mathcal{O}(1)$ as it is a constant.
\end{proof}

\begin{theorem}\label{thm:guarantee}

The gap between the surrogate immediate reward in Eq.\eqref{eq:surrogate_reward} and the original reward in Eq.\eqref{eq:reward} is bounded by $\max\{(q-(1-q)^{b-1})(\delta I_0-\delta I_{b-1}),(1-q-q^{b-1})(\delta I_0-\delta I_{b-1}) \}$.
\end{theorem}
\begin{proof}
The worst case happens when 1) for all $b'<b-1$, there is $\delta I(G,X_{t,b},a_{t,b'}) = \delta I_0$, or 2) for all $b'>1$, there is 
$\delta I(G,X_{t,b},a_{t,b'})=\delta I_{b-1}$. 
In case 1), the gap  between the exact expected marginal influence and our designed approximated one is:
\begin{align*}
&\quad \sum_{\beta=0}^{b-1} q^{\beta}(1-q)^{b-1-\beta} {b-1\choose \beta} [\delta I_0 - \delta I_{\beta}]\\
&\quad -(1-q)^{b-1}\delta I_0+(1-q)^{b-1}\delta I_{b-1}\\
&=\delta I_0-(1-q)\delta I_0 - q\delta I_{b-1} -(1-q^{b-1}(\delta I_0-\delta I_{b-1})\\
&=(q-(1-q)^{b-1})(\delta I_0-\delta I_{b-1})
\end{align*}

In case 2), the gap is:
\begin{align*}
&\quad \sum_{\beta=0}^{b-1} q^{\beta}(1-q)^{b-1-\beta} {b-1\choose \beta} [\delta I_{\beta} - \delta I_{b-1}]\\
&\quad -q^{b-1}\delta I_0+q^{b-1}\delta I_{b-1}\\
&=(1-q)\delta I_0+q \delta I_{b-1} -\delta I_{b-1} -q^{b-1}(\delta I_0-\delta I_{b-1})\\
&=(1-q-q^{b-1})(\delta I_0-\delta I_{b-1})    
\end{align*}
The bound is then $\max\{(q-(1-q)^{b-1})(\delta I_0-\delta I_{b-1}),(1-q-q^{b-1})(\delta I_0-\delta I_{b-1}) \}$.

\end{proof}

Theorem~\ref{thm:complexity} is critical as it makes the calculation of expected reward in our setting feasible. Meanwhile, Theorem~\ref{thm:guarantee} provides a guarantee to the approximation. It is worth noting that though the bound could be arbitrarily bad when $q\rightarrow 0$ or $q\rightarrow 1$, in practice this is rare. Moreover, the worst cases described in the proof are very extreme cases.
Empirical results show that even when $q = 0.2$ or $0.8$, RL4IM still practically works well. 

\begin{figure*}[ht]
    \centering
    \includegraphics[width=1\linewidth]{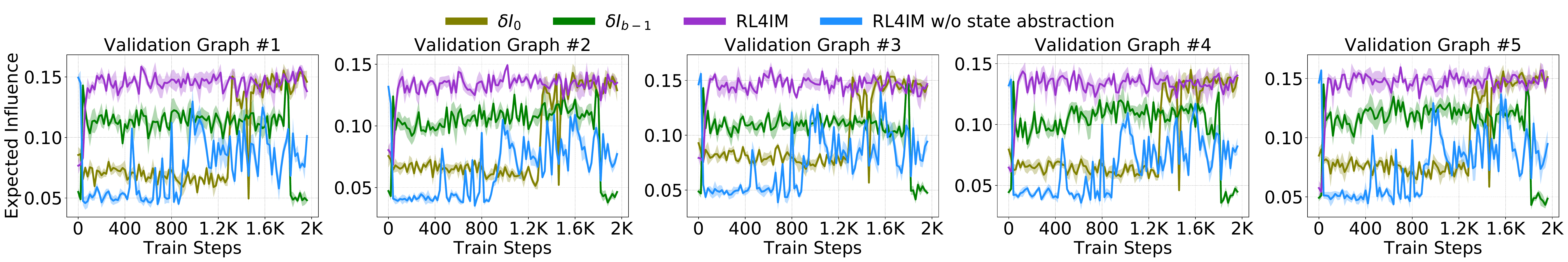}
    \caption{Validation curve on the 5 validation graphs during training for Q1. Shaded area indicates one standard deviation.}
    \label{fig:ablation}
\end{figure*}

\begin{figure*}
    \centering
    \includegraphics[width=1\linewidth]{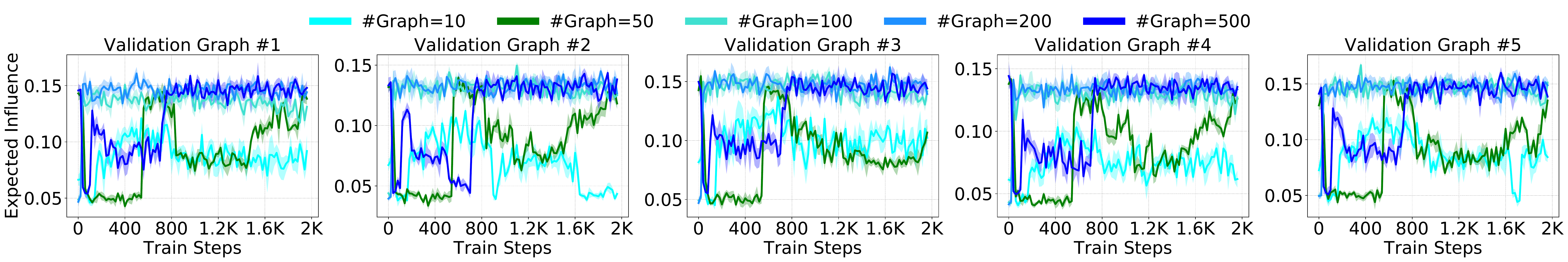}
   \caption{Validation curve on the 5 validation graphs for Q2. Shaded area indicates one standard deviation.}
    \label{fig:train_graph_number}
\end{figure*}

\section{Experiment}

\subsection{Experiment Settings}

\noindent\textbf{Environment} 
To evaluate the performances of different methods, we generate synthetic graphs using the powerlaw graphs \citep{onnela2007structure}, which is the Barabási–Albert (BA) growth model with an extra step that each random edge is followed by a chance of making an edge to one of its neighbors too. The average degree of a node is set to 3. The probability of adding a triangle after adding a random edge is set to 0.05. We will vary 1) the number of training graphs, 2) the willingness probability $q$, 3) intervention rounds $T$ and the per-round budget $B$, 4) the graph sizes $|V|$, and evaluate different methods under these varied settings. The belief propagation probability is of IM is set to 0.1. To get an influence number, the IM simulator runs 100 times and returns an average. All experiments are run on a Dell DSS 8440 Cauldron node, with a virtual environment with 2 Intel Xeon Gold 6148 2.4G CPU cores, 5G RAM, 1 NVIDIA Tesla V100 32G GPU, EDR Infiniband.
 
\noindent\textbf{Baselines} The baselines include 1) a greedy algorithm that adaptively selects seeds based on observation of the willingness status of selected nodes in previous rounds. It is part of the CHANGE algorithm \citep{wilder2018end} that is used in HIV prevention with node willingness uncertainty. 
2) S2V-DQN-IM which is adapted from S2V-DQN \cite{khalil2017learning} that combines RL with graph embedding. Note that this is the underlying architecture of recent works on RL for IM~\citep{li2019disco,tian2020deep}. We have added the state-abstraction component to it as we will show that the version without it barely converges well. The major difference between S2V-DQN-IM and RL4IM is it does not use our reward shaping technique, but estimates the reward using $\delta I_{b-1}$, i.e., assuming no uncertainty in a node's willingness.
3) Random which chooses nodes randomly.

\noindent\textbf{Evaluation setting}
In evaluation, we first generate a set of training graphs, and then generate another set of 5 held-out graphs that are used as validation set. The validation process will be activated approximately every 20 time steps (i.e., a checkpoint). During validation, it will run 20 episodes for each graph. The averaged reward over $20 \times 5 =100$ runs will be used as the metric to select the best hyperparameters as well as the model at the best checkpoint. The model selected using the validation set will then be evaluated in the test phase. During testing, 10 graphs will be generated that are unseen either in training or validation graphs. Each method will be run 20 times on a graph, totalling $20\times 10=200$ runs for one problem setting. 

For both S2V-DQN-IM and RL4IM, the following parameters are set to the same: memory size is 4096, 2) batch size is 32, 3) maximal training time steps is set to 2000, 4) the discount factor is 0.99, 5) the q-networks is an S2V-based graph embedding layer followed by a 128-neuron MLP layer, 6) the optimizer is Adam~\citep{kingma2015adam}. The other parameters, such as learning rate, exploration rate $\varepsilon$ and its decay rate are optimized from the validation set.

\subsection{Results \& Discussions}

The following values are set to default unless being evaluated: $|V|=200$, $T=2, B=4$, \#training graphs = 200; $q=0.6$. We are interested in the following questions.

\noindent\textbf{Q1: How does each new component of RL4IM affect the performance?}
Figure \ref{fig:ablation} shows the ablation study results. By removing either state-abstraction or our proposed reward shaping technique, the RL training becomes very unstable, or converges at a sub-optimal point. If we remove state-abstraction, then the state is very sparsely represented, and thus makes it hard for the graph embedding layer to effectively learn the optimal weights. On the other hand, by using either $\delta I_0$ or $\delta I_{b-1}$ as the reward, it leads to either over or under-estimation (which is the practice of existing works on RL for IM~\citep{li2019disco, tian2020deep,manchanda2020gcomb} that do not consider node uncertainty.

\noindent\textbf{Q2: Does the number of training graphs affect training performance?} 
To see whether and how the number of training graphs affect the performance, we show the validation curve during training, as in Figure~\ref{fig:train_graph_number}. It shows that by increasing the number of training graphs, the validation curve tends to converge at a more stable point. This is because with larger training graph pool, the RL agent is exposed to more environments and thus tends to generalize better among unseen graphs.
Based on this set of experiments, from now on all the experiments use 200 training graphs.

\begin{figure*}
    \centering
    \subfigure[vary $q$]{
        \centering
        \includegraphics[width=0.31\linewidth]{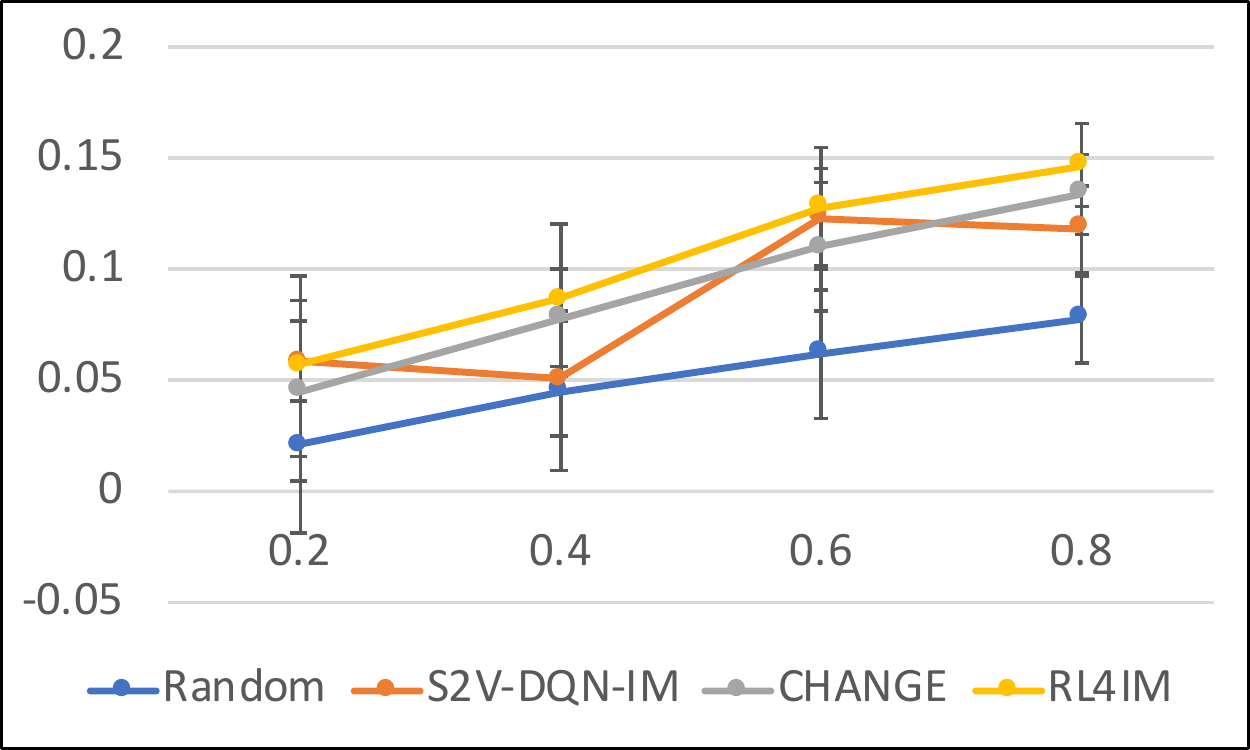}   \label{fig:vary_q}}
    \subfigure[vary $T$]{
        \centering
        \includegraphics[width=0.31\linewidth]{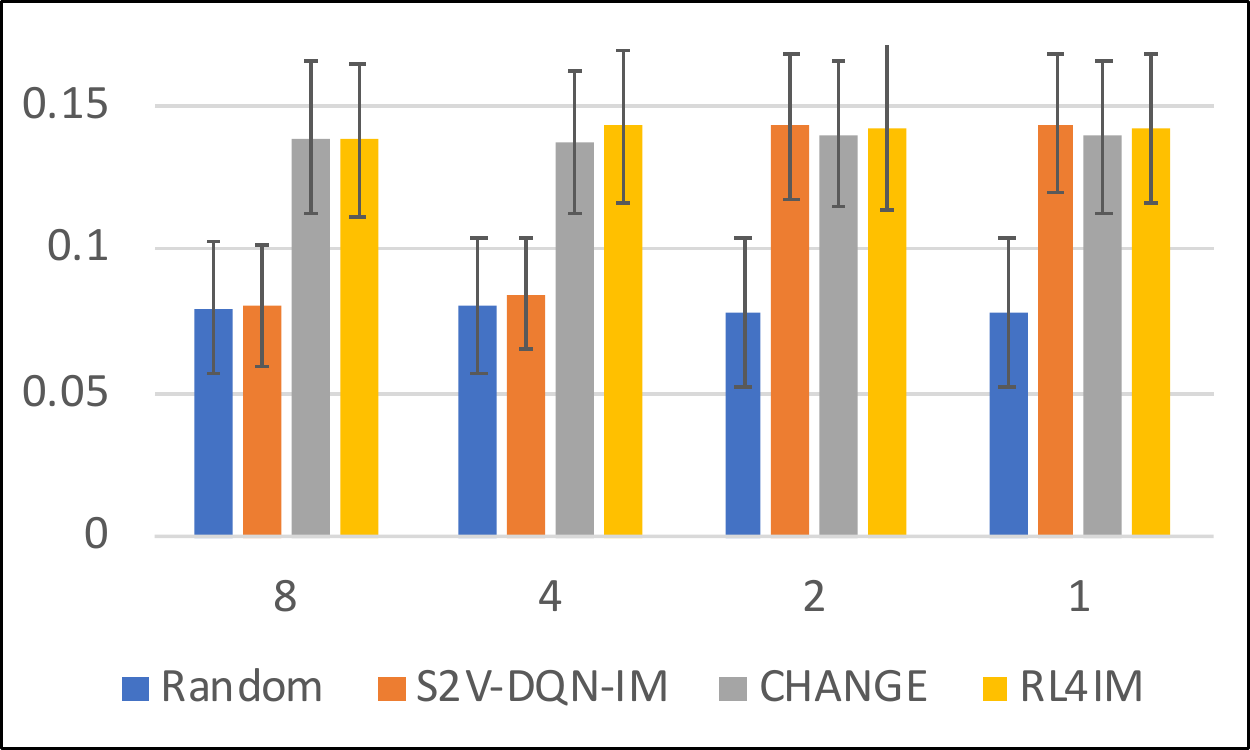}   \label{fig:vary_T}}
    \subfigure[vary $|V|$]{
        \centering
        \includegraphics[width=0.31\linewidth]{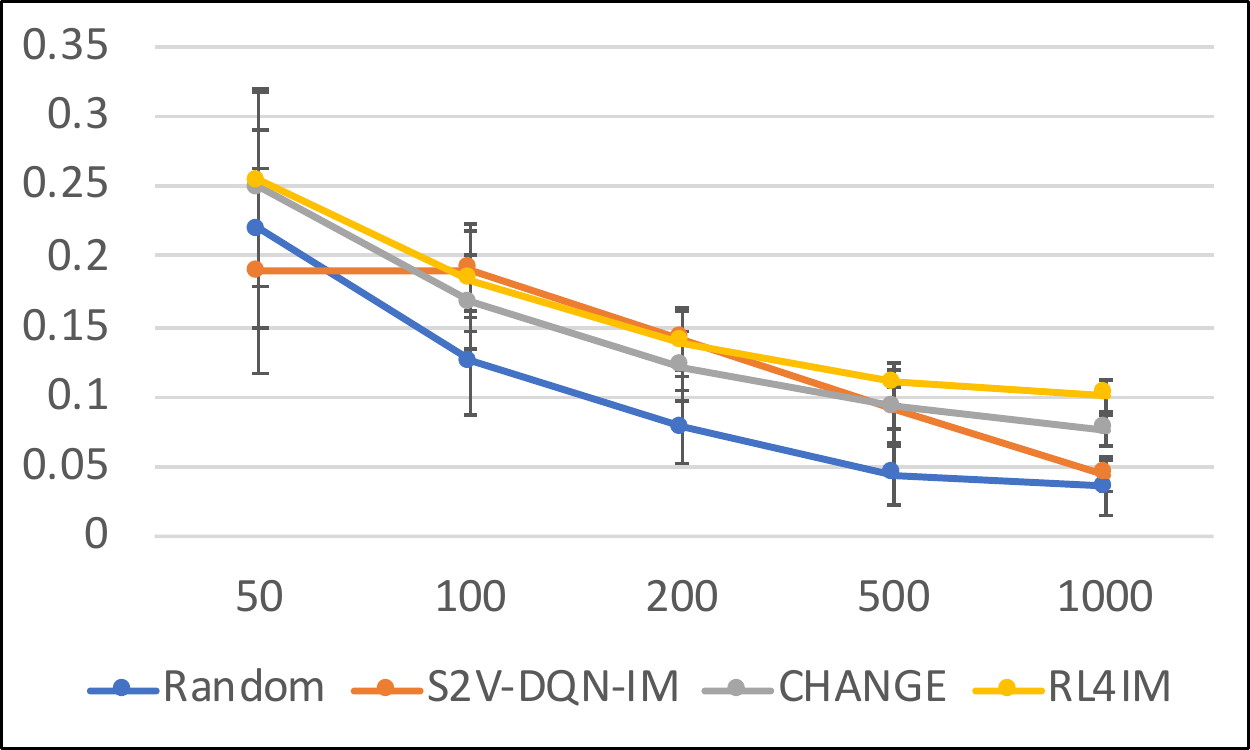}   \label{fig:vary_V}}
        
    \caption{Performance of different methods on unseen test graphs. The x-axis is the value of the underlying setting, and the y-axis is the expected normalized influence (w.r.t. the number of nodes) averaged over $10\times 20=200$ runs.}
    \label{fig:various_settings}
\end{figure*}

\noindent\textbf{Q3: Does RL4IM work well for different uncertainty values $q$?} Figure~\ref{fig:vary_q} shows that when $q$ increases, i.e., when nodes are more likely to be seeds when being selected, the expected influence grows higher. This is intuitive as the expected number of seed nodes becomes larger w.r.t. larger $q$ values.
Both RL4IM and CHANGE are better and more stable across different $q$ values, which are approximately twice the values of Random. The performance of S2V-DQN-IM is unstable as it uses a reward function that is far from the expected ground truth value.

\noindent\textbf{Q4: What if we decrease $T$ until 1?} In this setting, we fix $T\times B = 8$, and evaluate $T= 8, 4, 2, 1$. Note that when $T=1$, it is essentially a single round IM problem. Figure~\ref{fig:vary_T} shows the comparison results. This shows that RL4IM works well for single round IM problem as well. Similarly, CHANGE and RL4IM are the two best methods, while S2V-DQN-IM appears unstable at different settings.

\noindent\textbf{Q5: How does performance of RL4IM vary across different graph sizes?}  In this set of experiments we vary graph sizes whtin [50, 100, 200, 500, 1000]. Figure~\ref{fig:vary_V} shows that the normalized influence value decreases when the number of nodes $|V|$ increases. This is because the influence budget is fixed as $2\times 4=8$. When $|V|$ increases, the portion of nodes getting influence decreases. Similarly, CHANGE and RL4IM perform the best among all methods. 

From questions 3-5, the key takeaway is that RL4IM works robustly across different settings.
RL4IM is slightly better than CHANGE. This is potentially because RL4IM considers proactively the unwillingness of a node in its reward shaping component, whereas CHANGE only reactively adapts to realizations of the willingness status of nodes. Nonetheless, our main argument is that RL4IM, while performing as good as CHANGE, uses negligible runtime during test phase, and is therefore the better alternative for scenarios with low-resource computing. For example, once being trained, the policies are returned by RL within seconds even when the RL policy is run on a normal laptop.

\section{Conclusion}

We study the contingency-aware IM problem where a node's willingness to be a seed is uncertain. The state-of-the-art uses greedy algorithms to address the problem, but its slow run times are a barrier to transitioning this approach to low-resource settings as with non-profits serving marginalized populations. We propose a new learning-based perspective of solution to this problem using RL, so that it can output a seed selection strategy on a laptop within seconds at test time. Our major technical innovation is a theoretically grounded new algorithm, RL4IM, that exploits the properties of the underlying problem. Empirical results show that it matches the influence spread of the state-of-the-art, while having the advantage of negligible runtime during the test phase, a feature that is critical to low-resource non-profits. Our work is an example of RL for social good. We hope to shed some light to broader research areas in the low-resource computing paradigm.

\begin{acknowledgements} 
This work was supported by the Army Research Office (MURI W911NF1810208). Chen was supported by the Center for Research on Computation and Society.
\end{acknowledgements}

\bibliography{rl4im}

\end{document}